\documentclass[aip, apl,reprint]{revtex4-1}

\usepackage{amsthm}
\usepackage{amsmath}
\usepackage{bbm, dsfont}
\usepackage{graphicx}
\usepackage[colorlinks=true,citecolor=blue,urlcolor=black,linkcolor=blue]{hyperref}
\usepackage[table]{xcolor}
\usepackage{booktabs}

\newcommand{\be}{\begin{equation}}
\newcommand{\ee}{\end{equation}}

\newcommand{\del}[1]{}

\begin{document}
\title{Intrinsically-limited timing jitter in molybdenum silicide superconducting nanowire single-photon detectors} 
\author{Misael Caloz}\email{misael.caloz@unige.ch}
\affiliation{Group of Applied Physics, University of Geneva, CH-1211 Geneva, Switzerland}
\author{Boris Korzh}
\author{Edward Ramirez}
\affiliation{Jet Propulsion Laboratory, California Institute of Technology, Pasadena, CA 91109 USA}
\author{Christian Sch\"onenberger}
\affiliation{Department of Physics, University of Basel, CH-4056 Basel, Switzerland}
\author{Richard J.~Warburton}
\affiliation{Department of Physics, University of Basel, CH-4056 Basel, Switzerland}
\author{Hugo Zbinden}
\affiliation{Group of Applied Physics, University of Geneva, CH-1211 Geneva, Switzerland}
\author{Matthew D. Shaw} 
\affiliation{Jet Propulsion Laboratory, California Institute of Technology, Pasadena, CA 91109 USA}
\author{F\'elix Bussi\`eres} 
\affiliation{Group of Applied Physics, University of Geneva, CH-1211 Geneva, Switzerland}

\begin{abstract}

Recent progress in the development of superconducting nanowire single-photon detectors (SNSPDs) has delivered excellent performances, and has had a great impact on a range of research fields. The timing jitter, which denotes the temporal resolution of the detection, is a crucial parameter for many applications. Despite extensive work since their apparition, the lowest jitter achievable with SNSPDs is still not clear, and the origin of the intrinsic limits is not fully understood. Understanding its intrinsic behaviour and limits is a mandatory step toward improvements. Here, we report our experimental study on the intrinsically-limited timing jitter in molybdenum silicide (MoSi) SNSPDs. We show that to reach intrinsic jitter, several detector properties such as the latching current and the kinetic inductance of the devices have to be understood. The dependence on the nanowire cross-section and the energy dependence of the intrinsic jitter are exhibited, and the origin of the limits are explicited. System timing jitter of 6.0~ps at 532~nm and 10.6~ps at 1550~nm photon wavelength have been obtained.

\end{abstract}
\maketitle

Since their first demonstration~\cite{Goltsman2001a}, superconducting nanowire single-photon detectors (SNSPDs) have emerged as a key technology for optical quantum information processing~\cite{Hadfield2009}. Their low dark count rate, fast response time, small jitter, and high efficiency favours their use in various demanding quantum optics applications such as quantum key distribution~\cite{Boaron18}, quantum networking~\cite{Bussieres2014}, device-independent quantum information processing~\cite{Shalm2015a}, deep-space optical communication~\cite{Shaw14}, IR-imaging~\cite{Allman2015a,Zhao2017}, and integration in photonic circuits~\cite{Sprengers2011,Rath2015,Ferrari2018}. 


One advance in the SNSPD field has been the introduction of amorphous superconductors such as tungsten silicide (WSi)~\cite{Marsili2013} and molybdenum silicide (MoSi)~\cite{korneeva2014,Verma2015,Caloz2017a,Caloz2018}. SNSPDs based on these materials currently have the highest reported system detection efficiencies (SDE) (93\% for WSi~\cite{Marsili2013}). Their amorphous properties makes them materials of choice for applications where the film quality and yield are crucial, such as multi-mode coupled SNSPDs, or large arrays~\cite{Allmaras2017}.


The jitter is a crucial characteristic for time-resolved measurements such as light detection and ranging,  high-speed quantum communication, and lifetime measurement of single-photon sources. It denotes the timing variation of the arrival time of the detection pulses. Assuming independent contributions~\cite{You2013,Calandri2016} the total measured jitter can be written as the following:\\ $j_{system}^2 = j_{setup}^2 + j_{noise}^2 + j_{intrinsic}^2 + j_{geometric}^2$, where $j_{setup}$ includes the laser pulse width and measurement imprecisions, $j_{noise}$ is the contribution from the amplification and electronic parts, $ j_{intrinsic}$ includes the timing variation of the hotspot itself, and $ j_{geometric}$ is linked to the path the signal has to propagate depending on the photon absorption location in the nanowire~\cite{Calandri2016}. 

A wide range of values have been reported for different geometries and materials, typically from few to hundreds of picoseconds~\cite{Wu2017,You2013,Shcheslavskiy2016,Verma2015}. A record value of 2.7~ps at 400~nm wavelength have been recently achieved with a NbN device~\cite{Korzh2018}, which demonstrates that there is still room for improvements. Recent study with MoSi meandered devices showed that low jitter value (26~ps at 1550~nm) is achievable with amorphous material~\cite{Caloz2018}. Despite recent theoretical studies~\cite{Sidorova2017,Vodolazov2019}, the lowest experimental jitter achievable with SNSPDs is still not clear, and the fundamental limits remain unknown. Understanding its intrinsic behaviour and limits is a mandatory step toward improvements. This letter addresses this question with MoSi-based devices.


We report on devices with a special design which reduce significantly the geometric jitter component. They have been fabricated and measured with cryogenic amplifiers and measurement setup that reduces the noise jitter component~\cite{Korzh2018}. We show first that to reach intrinsically-limited jitter, different parameters have to be understood and optimized, namely the kinetic inductance ($L_k$) of the devices and its latching current ($I_{latch}$). Secondly, we tested devices with different cross-sections by varying their width and thickness to probe the nanowire cross-section dependence of the intrinsic jitter. Finally, we observed the photon-energy dependence of the jitter of MoSi SNSPDs.

\begin{figure}[t!] 
	\includegraphics[width=0.45\textwidth]{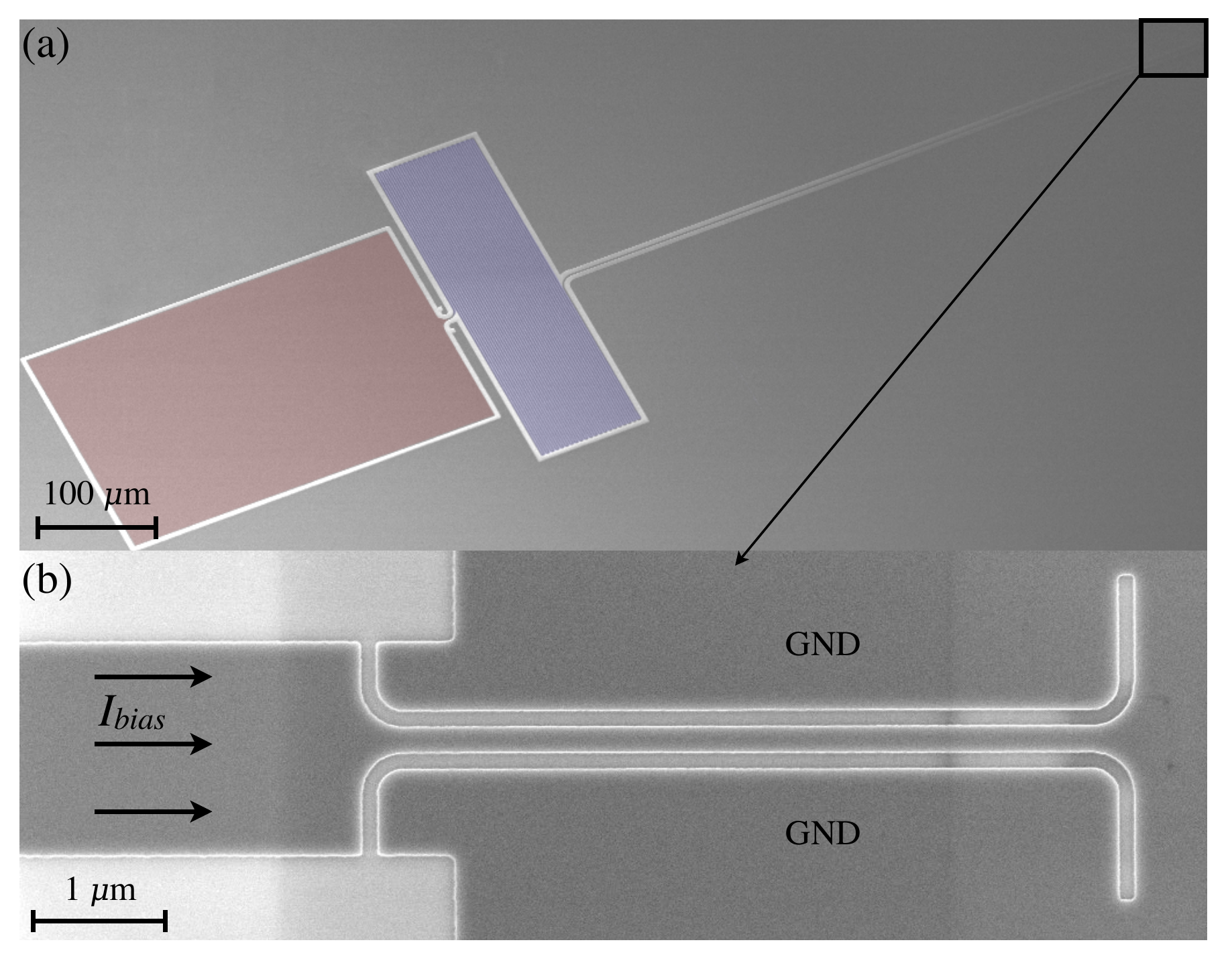}
	\caption{(a)~Scanning electron microscope image. The device is composed of a contact pad (in red), an inductor (in blue), and a nanowire connected to the ground. (b)~Zoom of the 5~$\mu$m long MoSi nanowire.}\label{fig1}
\end{figure}

The devices are fabricated out of 5, 7 and 9~nm-thick films of Mo$_{0.8}$Si$_{0.2}$ deposited by co-sputtering. The film is then patterned by a combination of e-beam lithography and reactive ion etching. A total of 80 different devices were measured for this study. The devices consist of a single 5~$\mu$m-long nanowire connected to a contact pad, through an meandered inductor, as illustrated in Fig.~\ref{fig1}. This nanowire design minimizes the geometric jitter component ($j_{geometric}$), while the series inductor is used to prevent the latching effect~\cite{Korzh2018}. To probe the nanowire cross-section dependence of the intrinsic jitter ($j_{intrinsic}$), the nanowire width is varied from 60~nm to 200~nm, depending on the thickness. When the cross-section of the nanowire increases, the bias current needed to operate the detector increases as well, and eventually gets larger than $I_{latch}$, which prevents its operation. To cope with this problem, the devices are tested with different series inductances ranging from 100~nH to 3500~nH.

The experiment was carried out using a pulse-tube cryocooler with a $\mathrm{^4He}$ sorption refrigerator reaching a base temperature just under 1~K. 

The signal from the SNSPD was amplified with SiGe cryogenic amplifiers from Cosmic Microwave. For the slew rate versus the kinetic inductance characterisation and the energy-dependence measurements, the CITLF3 and the CITLF132 were used, respectively. The SNSPDs were biased with a low-noise current source through a 5~k$\Omega$ resistive bias-T at the input of the AC-coupled amplifier. In order to get higher latching currents, a shunt inductance of 1.2~$\mu$H was connected to the ground through a 50~$\Omega$ resistance.  

To investigate the photon-energy dependence of the jitter, we used two second harmonic generation (SHG) crystals to frequency double the mode-locked lasers from 1064~nm and 1550~nm  to 532~nm and 775~nm, respectively. After the crystal, the light was collimated and free-space coupled into the cryostat through a series of glass windows in the vacuum chamber and the heat shields at 40~K and 4~K, flood illuminating the device under test. The optical intensity was controlled with a circular metallic variable neutral-density filter. This configuration ensured that the converted and unconverted light co-propagated via the same path through the optical setup. After generation, filters were used to select 532, 775, 1064, and 1550~nm wavelength illumination.

The SNSPDs and laser synchronization signals were acquired simultaneously on a digital real-time oscilloscope with a sampling rate of 40~GS/s and a bandwidth of 12~GHz. The time delay between the two pulses was recorded for each acquisition. Histograms of 5000~detection delays were collected for each jitter measurement, which typically required a collection time of approximately 5~minutes. The histogram are fitted with an exponentially modified gaussian (EMG) function, and the jitter was obtained by taking the FWHM of this distribution.

The first part of this work consisted in understanding the latching current and noise jitter dependence on the electronic readout and device kinetic inductance. The desired mode of operation for SNSPDs is achieved only when the electric feedback is slower than the nanowire cooling time, which happens naturally if its kinetic inductance is large enough~\cite{Kerman2009}. If this feedback is sped up by decreasing the kinetic inductance, the device will suffer from the latching effect where it is locked in a resistive state and can no longer detect photons. When the cross-section of the nanowire increases, \textit{i.e.}\ when either its width and/or thickness increases, the switching current $I_{sw}$ increases as well, and can eventually get bigger than $I_{latch}$. The large kinetic inductance is necessary to slow down the electric feedback and prevents latching. However, if $L_k$ is too large, two problems arise: (i) the maximum count rate of the SNSPD is reduced, and (ii) the electrical signal coming out of the nanowire after a detection is slowed down, meaning a lower slew rate ($SR$) and consequently a larger noise jitter. The last point is crucial to reach intrinsically-limited jitter. The jitter induced by the gaussian noise of the cryogenic amplifier can be estimated by~\cite{You2013,Caloz2018}: 
\begin{equation}\label{eq1}
j_{noise} = 2 \sqrt{2 \ln 2}  \frac{\sigma_\mathrm{RMS}}{SR}   
\end{equation}
where $\sigma_\mathrm{RMS}$ is the amplifier RMS noise and $SR$ is the slew rate of the detection signal.

\begin{figure}[t!] 
	\includegraphics[width=0.5\textwidth]{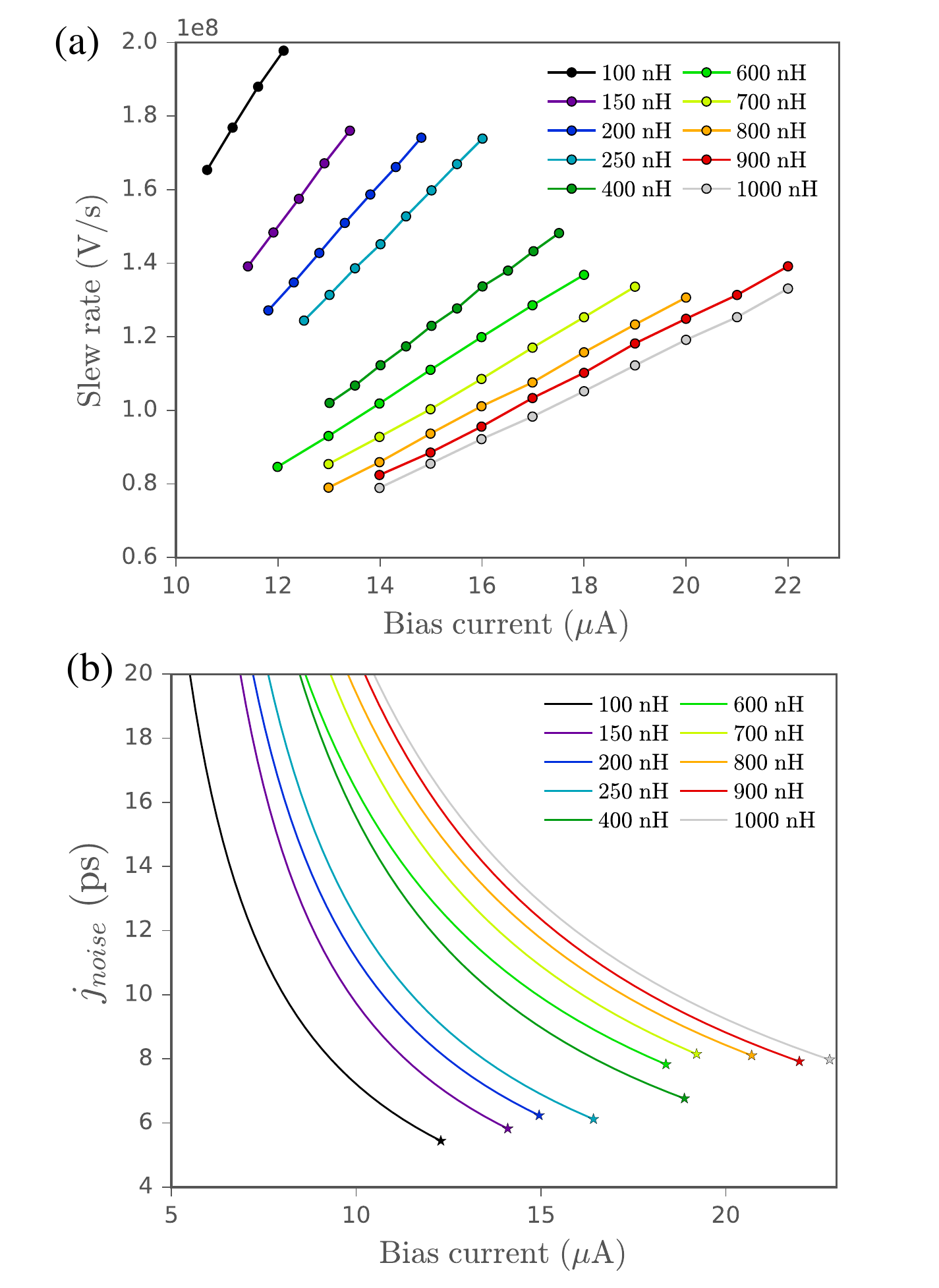}
	\caption{Dataset for 120~nm wide, 7~nm thick nanowires, with different kinetic inductance. (a) Slew rate of the signal rising edge for devices with different kinetic inductance (shown in the legend). (b) Estimated jitter induced by the electrical noise (described in Eq.~\ref{eq1}) as a function of the bias current. The stars indicate the latching current for the corresponding devices.}\label{fig2}
\end{figure}

Fig.~\ref{fig2} shows the slew rate and the jitter induced by the noise for 120~nm wide, 7~nm thick nanowires, with different kinetic inductance ranging from 100~nH to 1000~nH. The slew rate is extracted from oscilloscope traces and is plotted in Fig.~\ref{fig2}a as a function of the bias current. Fig.~\ref{fig2}b shows the corresponding estimated jitter induced by the electrical noise as described in Eq.~\ref{eq1} as a function of the bias current.

 A compromise between $L_k$, $I_{latch}$, $SR$ and consequently $j_{system}$ has to be found. Regarding Fig.~\ref{fig2}, it is clear that the best compromise is to reduce as much as possible $L_k$, while satisfying $I_{latch}>I_{sw}$. The lowest noise jitter we could achieve was estimated to be 5~ps. We performed this characterization for the three thicknesses, and obtained quantitatively the same results. The only way left to reduce the noise jitter is by decreasing the kinetic inductance of the device, which is incompatible with the latching effect, as explained above. 
 
Once the latching current for a given $L_k$ is known, we selected devices with the lowest $L_k$ possible that still satisfied $I_{latch}>I_{sw}$ to ensure optimal performances. The best detectors for each thickness were selected for energy-dependence measurements as shown in Tab.~\ref{tab2}. We experimentally observed higher latching currents when using the CITLF1 amplifier, this allowed us to pick lower kinetic inductances resulting in lower system jitter. This improvement is probably due to electrical reflections going back and forth between the SNSPD and the amplifier but the exact explanations is left for future work.

\begin{figure*}[t!] 
	\includegraphics[width=\textwidth]{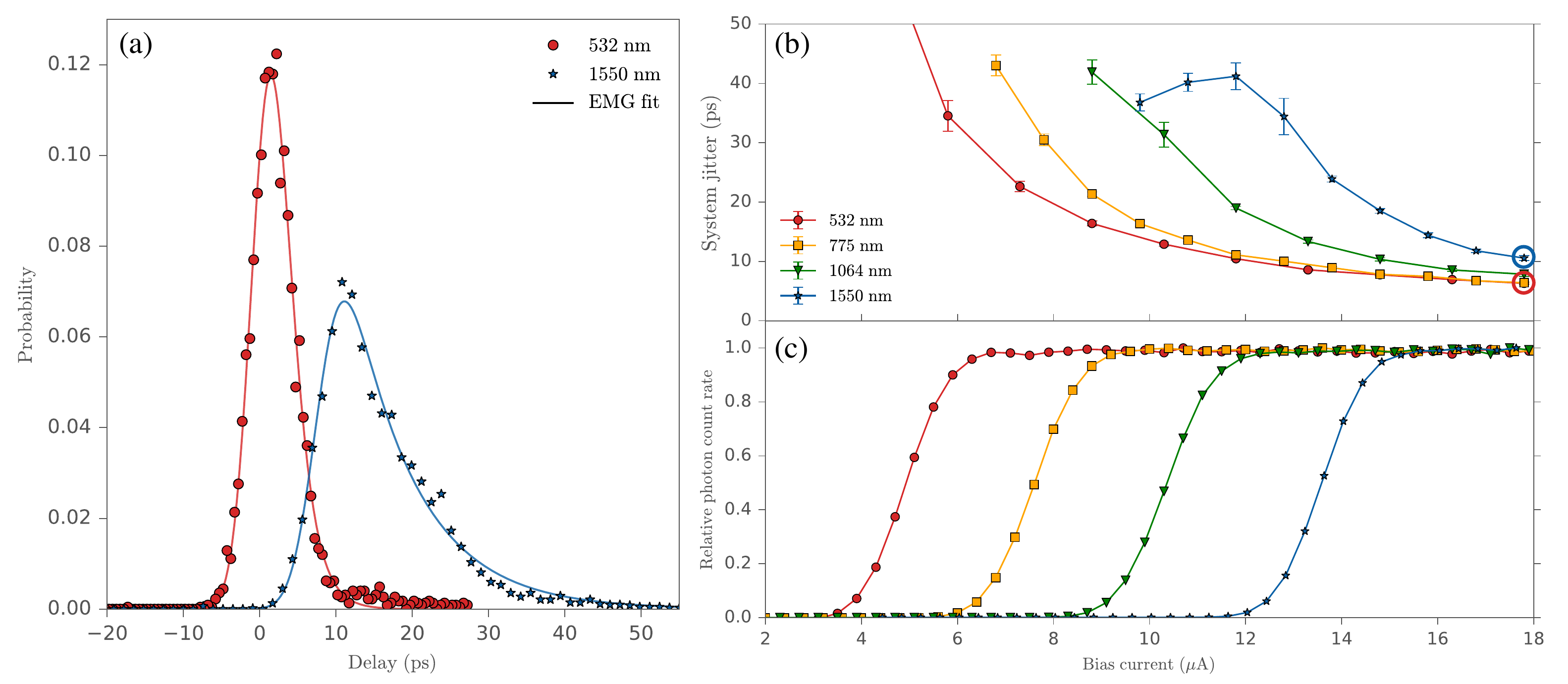}
	\caption{Results for 7~nm-thick and 100~nm-large device. (a)~Jitter histogram at 532 and 1550~nm for a bias current of $17.9\ \mu$A, as indicated by the circles in (b). The lines represent the Exponentially Modified Gaussian fit. (b)~Jitter FWHM as a function of the bias current, for different wavelengths. (c) Photon count rates for the same wavelengths as shown in (b).}\label{fig3}
\end{figure*}

Fig.~\ref{fig3}a shows the timing histogram for the 7~nm thick device measured for a bias current of $17.9\ \mu$A, as indicated by the circles in Fig.~\ref{fig3}b. The FWHM of the distributions are 6.0~$\pm$~0.2~ps and 10.6~$\pm$~0.2~ps at 532~nm and 1550~nm wavelength, respectively. A non-gaussian tail is clearly observed, it becomes more apparent for long wavelengths and low bias currents, this behaviour has also been reported in many studies~\cite{Sidorova2017,Caloz2018,Korzh2018}, but its origin remains unclear. The jitter energy dependence is plotted in Fig.~\ref{fig3}b. The 5~nm and 9~nm-thick devices exhibit qualitatively the same behaviour, the results are summarized in Tab.~\ref{tab2}. One notable difference between cross-sections is that we could obtain a jitter of 14.5~ps at 1550~nm for the 9~nm-thick device, while the 5 and 7~nm-thick devices showed values close to 10~ps. 

One clear signature of the intrinsic jitter can be seen in Fig.\,\ref{fig3}b: for the same bias current, $j_{setup}$, $j_{noise}$, and $j_{geometric}$ are independent of the photon wavelength. This means that the difference between different wavelengths is only due to the $j_{intrinsic}$ and that we are probing the intrinsic component. The intrinsic behaviour becomes clearer for low bias currents and longer wavelengths.

The noise jitter of device \#2 was estimated to be 5~ps. By subtracting it to its system jitter of 6.0~ps, \textit{i.e.}\ the noise contribution would be totally negligible, the remaining jitter goes down to roughly 3.5~ps. While this value seems very close to 2.7~ps of the NbN-based devices~\cite{Korzh2018}, the difference appears more clearly for longer wavelengths, where NbN achieved 4.6~ps at 1550~nm, while MoSi achieved 8.6~ps (after subtracting the noise). This shows that MoSi-based devices seem to yield larger jitter and the fact that this remaining jitter is purely intrinsic seems to point to a material difference for long wavelengths.

The energy-dependence data sets are currently being analysed following a recent theoretical framework including the intrinsic jitter behaviour and Fano fluctuations~\cite{Allmaras2018}. While this model fits very well data with NbN devices, our study adds experimental inputs and could lead to a better understanding of intrinsic jitter mechanism and to an unified detection mechanism model. 

\begin{table}[t!] 
	\caption{List of the selected devices for energy-dependence measurements, and their system jitter.}
	\label{tab2}
	\begin{ruledtabular}
		\begin{tabular}{cccc|llll}
			&Thickness & Width& $L_k$ & \multicolumn{4}{c}{Wavelength (nm)}  \\
			&(nm) & (nm) & (nH) & 532 & 775 & 1064 & 1550  \\
			\hline
			\#1  & 5		& 120 & 200 & 6.2 & 6.5 & 8.8 & 10.7~ps    \\ 
			\#2 &7		   & 100 & 200 & 6.0 & 6.4 & 7.8 & 10.6~ps     \\  
			\#3 &9  	   & 80   & 250 & 7.0 & 7.3 & 9.5 & 14.4~ps    \\ 
			       
		\end{tabular}
	\end{ruledtabular}
\end{table}



In conclusion, we reported intrinsically-limited timing jitter with MoSi SNSPDs. To reach fundamental limits, we showed that the kinetic inductance has to be minimized taking into account the latching current of the detector. We developed an experimental setup that minimize every component of the system jitter of our SNSPDs and allows us to probe and quantify the intrinsic jitter. The energy-dependence of the intrinsic jitter is shown and points to a material limitation of MoSi-based devices for wavelengths longer than 1064~nm. Finally, we observed that the intrinsic jitter is higher for thicker devices and longer wavelengths.



The authors would like to acknowledge the Swiss NCCR QSIT (National Center of Competence in Research - Quantum Science and Technology) for financial support. Part of this work was conducted at the Jet Propulsion Laboratory, California Institute of Technology, under contract with the National Aeronautics and Space Administration.






%






\end{document}